\documentclass[aps,prl,twocolumn,superscriptaddress,showpacs,longbibliography]{revtex4-2}
\usepackage{amssymb}
\usepackage{graphicx}
\usepackage{dcolumn}
\usepackage{bm}
\usepackage{amsmath}
\usepackage{subfigure}
\usepackage{float}
\usepackage{color}
\usepackage{xcolor}
\usepackage[colorlinks=true,citecolor=blue]{hyperref}

\usepackage{braket}
\usepackage{indentfirst}
\usepackage{newtxtext}
\usepackage{booktabs}

\begin{document}

\title{Origin and classification of fractional dynamical topology in non-Hermitian quantum systems}

\author{Zimiao Zhang}
\affiliation{College of Physics and Electronic Engineering, Shanxi University,  Shanxi University, Taiyuan, Shanxi 030006, China}

\author{Li Chen}
\email{lchen@sxu.edu.cn}
\affiliation{College of Physics and Electronic Engineering, Shanxi University,  Shanxi University, Taiyuan, Shanxi 030006, China}
\affiliation{State Key Laboratory of Quantum Optics Technologies and Devices,
Institute of Theoretical Physics, Shanxi University, Taiyuan 030006, China}

\author{Weilun Jiang}
\email{wljiang@sxu.edu.cn}
\affiliation{State Key Laboratory of Quantum Optics Technologies and Devices, Institute of Opto-electronics, Shanxi University, Taiyuan, Shanxi 030006, China}
\affiliation{Collaborative Innovation Center of Extreme Optics, Shanxi University, Taiyuan, Shanxi 030006, China}
\affiliation{College of Physics and Electronic Engineering, Shanxi University,  Shanxi University, Taiyuan, Shanxi 030006, China}

\author{Heng Shen}
\affiliation{State Key Laboratory of Quantum Optics Technologies and Devices, Institute of Opto-electronics, Shanxi University, Taiyuan, Shanxi 030006, China}
\affiliation{Collaborative Innovation Center of Extreme Optics, Shanxi University, Taiyuan, Shanxi 030006, China}
\affiliation{College of Physics and Electronic Engineering, Shanxi University,  Shanxi University, Taiyuan, Shanxi 030006, China}

\begin{abstract}

Dynamical quantum phase transitions (DQPT) provide a fundamental framework for characterizing nonequilibrium critical phenomena, and have recently been extended to non-Hermitian systems. Half-integer jumps of the dynamical topological order parameter can arise in both Hermitian and non-Hermitian systems, yet their origin in non-Hermitian settings remains unclear. Here, we investigate DQPTs in a two-band non-Hermitian model and show that these fractional jumps originate from the discontinuity in the complex energy spectrum. We demonstrate that this behavior is intrinsically linked to the steady-state phase diagram of the quench Hamiltonian, and also classify the occurrence of DQPT. We further extend our analysis to the multiband Hamiltonian and the two-dimensional model, providing a unified understanding of fractional dynamical topological signatures.

\end{abstract}

\maketitle

\textcolor{blue}{\it Introduction.--} 
Dynamical quantum phase transitions (DQPTs) have emerged as a powerful concept in non-equilibrium quantum many-body physics, serving as the dynamical counterpart of equilibrium phase transitions \cite{heyl2013dynamical,heyl2015scaling,heyl2018dynamical,zhang2017observation,flaschner2018observation,Guo2019Observation,Kai2020Probing,Nie2020Experimental,sharma2016slow,Bhattacharya2017Mixed,yang2019floquet,Peotta2021Determination,jafari2021floquet,jafari2022floquet,Zeng2023Dynamical}. Following a quantum quench, DQPTs manifest as nonanalyticities in the Loschmidt rate (LR) at critical times $t_c$, accompanied by a sudden jump of the dynamical topological order parameter (DTOP) \cite{budich2016dynamical}. Specifically, in Hermitian systems, the DTOP has been extensively studied, where integer jumps are commonly observed \cite{jurcevic2017direct,Wang2019Simulating,Tian2019Observation,xu2020measuring,Chen2020Experimentally,Heyl2017Quenching,Lang2018Dynamical,Sadrzadeh2021Dynamical,Yu2021Correlations,naji2022dissipative}.
With continuous breakthroughs in non-Hermitian theory and corresponding advances in experimental systems \cite{bender1998real,Non-Hermitian2020Yuto,bergholtz2021exceptional,gong2018topological,shen2018topological,kawabata2019symmetry,yin2018geometrical,yao2018edge,yokomizo2019non,ghatak2019new,helbig2020generalized,Zhang2020Correspondence,yokomizo2020non,Xiao2020Non-Hermitian,wang2021topological,zhang2021observation,ding2022non}, the concept of DTOP has been extended to non-Hermitian systems, where half-integer jumps have also been reported \cite{Zhou2021Non-Hermitian,tang2022dynamical,sun2022biorthogonal,mondal2022anomaly,Mondal2023Finite,jing2024biorthogonal,Zhang2025Self-normal,ZHANG2026131193,Xu_2026}. 

Despite these advances, the physical origin of half-integer jumps remains largely unclear. Several key questions remain unanswered: what underlying mechanism gives rise to fractional quantization of the DTOP? How is it related to the other dynamical observables? Moreover, the behavior of the half-integer DTOP and its relationship with the steady state phase diagram has not been systematically clarified. Here, we answer this question by investigating DQPTs in a two-band non-Hermitian model. By employing both the  biorthogonal and self-normalized basis \cite{PhysRevLett.121.026808,edvardsson2019non,edvardsson2020phase}, we demonstrate that the emergence of half-integer DTOP jumps originates from the energy discontinuity, and is intrinsically linked to the steady state phase diagram of the quench Hamiltonian. We proved that this result can be generalized to multiband Hamiltonians, although with certain restrictions on their form. Furthermore, we extend it to the two-dimensional Haldane model, and also identify fractional jump under the analog of DTOP. Our results offer a unified understanding of fractional dynamical topological signatures and provide experimentally accessible criteria for probing critical dynamics in non-Hermitian quantum systems.

\begin{figure}[t]
\centering
\includegraphics[width=0.47\textwidth]{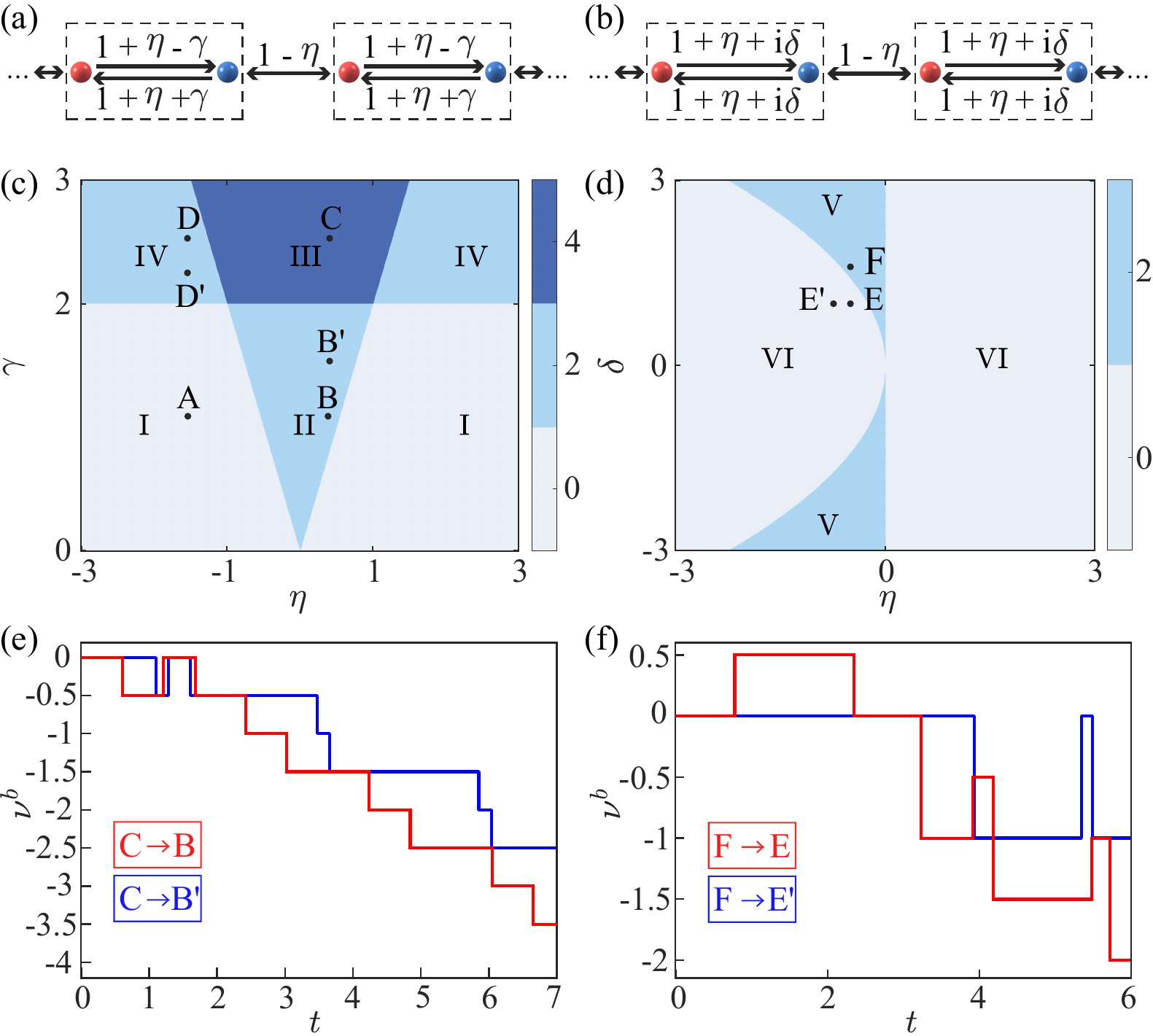}
\caption{\label{fig1} (a,b) Sketch map of $(\eta,\gamma) $ and $(\eta,\delta)$ model. The dashed boxes indicate the unit cell, with two sublattices colored by red and blue. (c,d) The phase diagrams in the $(\eta,\gamma)$ and $(\eta,\delta)$ planes. The color depth is determined by the number of momentum point at which the real part of the 
energy vanishes \cite{SM}. From parameter points A to F, the associated coordinates are (-1.4,1), (0.4,1), (0.4,1.5), (0.4,2.5), (-1.4,2.5), (-1.4,2.25), (-0.5,1), (-0.8,1), (-0.5,1.5). (e,f) Dynamics of Biorthogonal DTOP for two quench processes. Here, $H^i$s are the same while $H^f$s are close and in the same phase region. }
\end{figure}

\textcolor{blue}{\it Model.--} We consider one-dimensional non-Hermitian Su--Schrieffer--Heeger (SSH) model \cite{su1980soliton}, as shown in Fig.~\ref{fig1}(a) and (b). Here, $\eta$ characterizes the dimerization of the SSH model and distinguishes different topological phases in the Hermitian limit; $\gamma$ and $\delta$ denote non-reciprocal hopping processes and non-Hermitian correction to the intra-sublattice hopping amplitude, respectively. In momentum space, the non-interacting Hamiltonian is given by a two-band model
\begin{equation}
	H_k = d_k^x\,\sigma^x + d_k^y\,\sigma^y + d_k^z\,\sigma^z =  \mathbf{d}_k \cdot \boldsymbol{\sigma} ,
    \label{eq1}
\end{equation}
with the Bloch vector $\mathbf{d}_k = (d_k^x, d_k^y, d_k^z)$ and the vector of the Pauli matrices
$\boldsymbol{\sigma} = (\sigma^x, \sigma^y, \sigma^z)$. 
The coefficients $d_k^{\alpha}$ with $\alpha = x,y,z$ include both Hermitian and non-Hermitian contributions in terms of $d_k^x = (1+\eta) + (1-\eta)\cos k + i\delta$, $d_k^y = (1-\eta)\sin k + i\gamma$, and $d_k^z = 0$.

To facilitate analysis, we divide the phase region according to the number of the momentum point at which the real part of the 
energy vanishes. Note that the phase boundaries obtained by this method are consistent with those derived from the winding number, which corresponds to the emergence of exceptional points \cite{SM}.  In particular, we investigate $(\eta,\delta)$ and $(\eta,\gamma)$ phase diagrams simultaneously [Fig.~\ref{fig1}(c) and (d)]. This is due to the fact that those momentum points are independent of the Hamiltonian parameters for the former, whereas in the latter they exhibit explicit parameter dependence. To compare its influence on DTOP dynamics, we explore the dynamics of two adjacent quench processes in each phase diagram. The results indicate that half-integer jumps emerges for both processes in the $(\eta, \gamma)$ diagram [Fig.~\ref{fig1}(e)], while this behavior does not apply to the $(\eta, \delta)$ diagram [Fig.~\ref{fig1}(f)], which will be investigated in the following text.


\textcolor{blue}{\it Dynamical quantum phase transition.--} In order to fully describe the eigenstructure of a non-Hermitian Hamiltonian $H_k$, we use the right and left eigenstates, defined as \cite{brody2014biorthogonal}
\begin{equation}
	H_k\ket{\psi_{k,\alpha}^R}
	= \epsilon_{k,\alpha}\ket{\psi_{k,\alpha}^R},
	H_k^\dagger\ket{\psi_{k,\alpha}^L}
	= \epsilon_{k,\alpha}^*\ket{\psi_{k,\alpha}^L},
    \label{eq2}
\end{equation}
where $\epsilon_{k,\alpha} = \alpha \sqrt{(d_k^x)^2+(d_k^y)^2+(d_k^z)^2}$ is the energy at each momentum point and $\alpha=\pm$ labels the two bands. Here, the left and right eigenstates satisfy the biorthonormality $\braket{\psi_{k,\alpha}^L|\psi_{k',\beta}^R} = \delta_{kk'}\delta_{\alpha\beta} $ and completeness relations $\sum_{k,\alpha} \ket{\psi_{k,\alpha}^R}\bra{\psi_{k,\alpha}^L} = \mathbb{I}$. Now we consider a sudden quench of the system from $H^i$ to $H^f$, as depicted by the parameter points in Fig.~\ref{fig1} (c) and (d). The initial state is prepared by filling the lower band for all momentum $k$, corresponding to the steady state of $H^i$. The Loschmidt echo $\mathcal{L}^\mu$ is defined as the overlap between the initial state and the time-evolved state, satisfying $\mathcal{L}^\mu(t) = \prod_k C_k\left|g_k^\mu(t)\right|^2$, where $\mu=b,s$ labels the choice of biorthogonal or self-normalized basis and $C_k$ is the modulus of the corresponding time-evolved state \cite{SM}. Here, $g_k^\mu(t)$ is expressed as \cite{mondal2022anomaly,jing2024biorthogonal} 
\begin{equation}
	g_k^\mu(t) = \cos(\epsilon_{k,+}^ft) - i\sin(\epsilon_{k,+}^ft) \mathcal{F}_k^\mu,
    \label{eq3}
\end{equation}
where $\mathcal{F}_k^\mu$ is a function depending on the basis, 
\begin{equation}
	\mathcal{F}_k^b = -\frac{\mathbf{d}_k^f \cdot \mathbf{d}_k^i}{\epsilon_{k,+}^f  \epsilon_{k,+}^i}, \quad \mathcal{F}_k^s = \frac{\braket{\psi_{k,-}^{iR}|H_k^{f}|\psi_{k,-}^{iR}}}{\epsilon_{k,+}^f}.
    \label{eq4}
\end{equation}
Therein lies simplified expression in the frame of biorthogonal basis \cite{SM}. It is known that a zero value of $g_k^{\mu}(t)$ at $(k_c,t_c)$ defines a dynamical critical point, with $k_c$ the critical momentum. The Fisher zeros that happen at imaginary time $z = it$ can be obtained analytically as
\begin{equation}
	z_{k,n}^\mu = i\frac{(2n+1)\pi}{2\epsilon_{k,+}^f} + \frac{1}{\epsilon_{k,+}^f}\tanh^{-1} \mathcal{F}_k^\mu, \label{eq5}	
\end{equation}
where different branches are labeled by an integer $n$.

In order to demonstrate the DQPT, we exploit the LR, defined as $\lambda^{\mu}(t) = - \lim_{N \to \infty} \frac{1}{N} \ln \mathcal{L}^{\mu}(t)$, where $N$ is the system size. To avoid inaccurate numerical integration on the calculations of the conventional geometrical phase \cite{zhou2018dynamical,gong2018piecewise,aharonov1987phase}, we define the DTOP by directly summing up the discontinuous jump of the phase $\phi_k^\mu(t) \equiv \arg[g_k^\mu(t)]$ occurring exactly at the critical momentum, which can be written as
\begin{equation}
	\nu^{\mu}(t)
	= \frac{1}{2\pi}
	\sum_{k_j}
	\Delta \phi_{k_j}^{\mu}(t).
    \label{eq6}
\end{equation}
Here $k_j$ are the discontinuity points of the phase $\phi_{k}^\mu(t)$ in the effective Brillouin zone (EBZ). Note the result under this definition is consistent with the direct calculation on the geometrical phase \cite{SM}.

\textcolor{blue}{\it Description of non-Hermitian DQPT.--} We proceed to explore the origin of the half-integer jump through the above expressions, in comparison with integer jumps. In order to exhibit an intuitive picture, we analyze this quench protocol based on simplified formulas obtained in the biorthogonal framework. According to Eq.~\eqref{eq4}, although the numerator $\mathbf{d}_k^i\cdot\mathbf{d}_k^f$ is a continuous function of $k$, the energy spectrum $\epsilon_{k}$ could contribute to the sudden jump.

\begin{figure}
\centering
\includegraphics[width=0.47\textwidth]{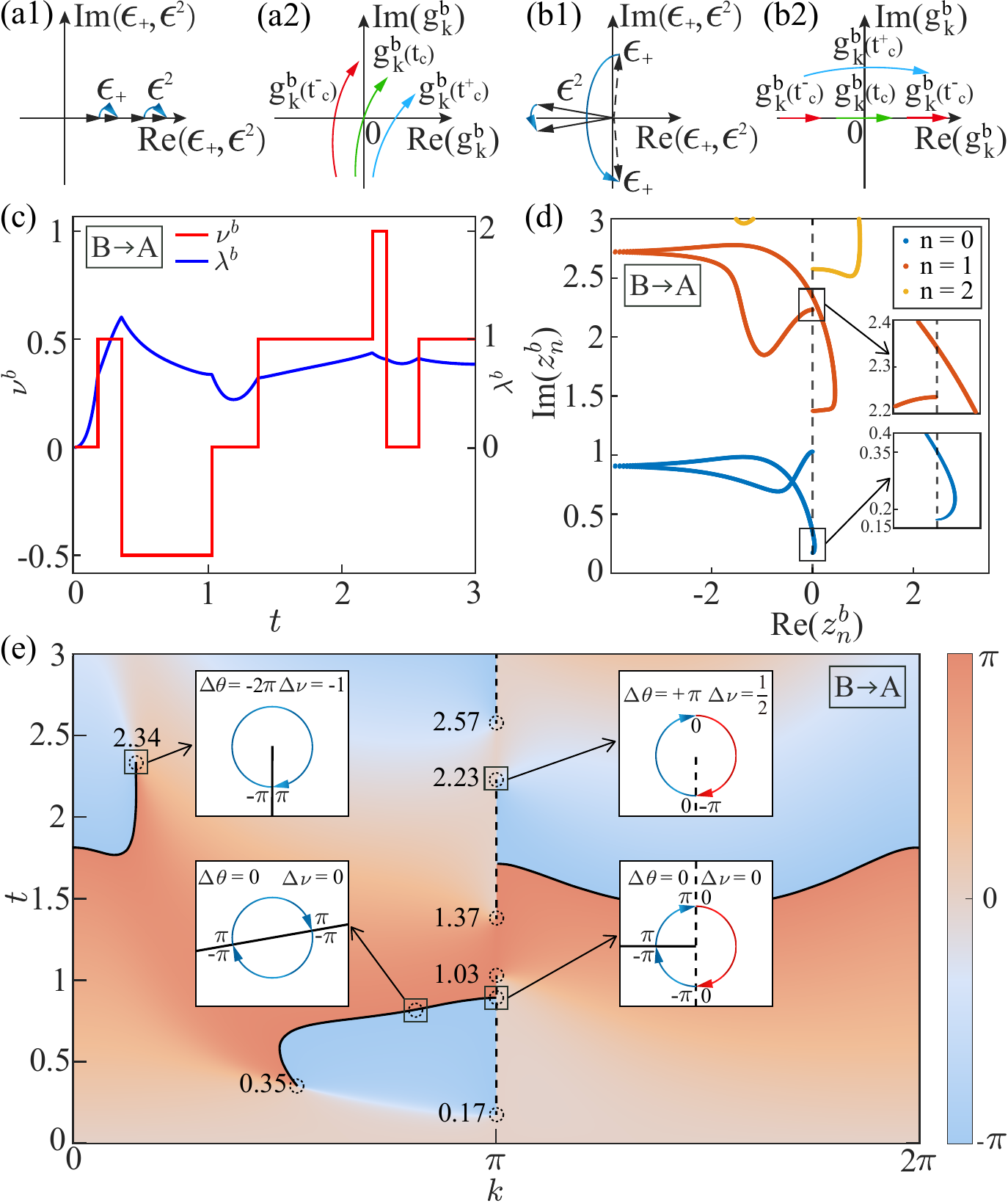}
\caption{\label{Fig2} Dynamical quantum phase transition for a sudden quench from B to A in the biorthogonal basis. The energy and phase in complex vector form for Hermitian case in (a) shows the integer jump of DTOP, and the non-Hermitian case in (b) shows the half-integer jump. The blue arrows in (a1) and (b1) represent the change between two $k$ values near $k_c$, while the dashed black arrow denotes the energy value. The colored arrows in (a2) and (b2) represent the continuous scan with respect to $k$ at different time near $t_c$. (c) Time-evolution of DTOP and LR. (d) The Fisher zeros in the complex-time plane.  The curves with different colors label the three branches of the Fisher zeros. The panels show the zoomed-in results near $t_c$. (e) The dynamical phase $\phi_k^{b}$ in $(k,t)$ plane. All $(k_c, t_c)$ points are marked with circles. The zoomed-in panels show the value of the winding numbers for four typical cases, the endpoints of the dashed and solid lines, as well as their interior points. }
\end{figure}

In the Hermitian case, $\epsilon_{k}$ is a real number  [Fig.~\ref{Fig2}(a1)], thereby $g_k^b(t)$ is a continuous function with respect to $k$ at a fixed time $t$. The phase jump $\Delta \phi_{k_j}^{b}(t)$ occurs when the its value, as a function of $k$, passes through the negative real axis, thus the phase $\phi_k^b(t)$ undergoes a discontinuous change between $\pi$ and $-\pi$ [Fig.~\ref{Fig2}(a2)], contributing an integer value to the DTOP. In particular, at $t=t_c$, the function curve $g_k^b(t)$ passes through the origin. In the vicinity of the critical point $t=t_c^-$ and $t=t_c^+$, one crosses the real axis from the left side, while the other crosses from the right side, finally leading to a jump of the function $\nu^b(t)$. This indicates that the change of DTOP is only associated with the winding number defined on $(k,t)$ plane around an isolated zero point. 

By contrast, in non-Hermitian systems, $\epsilon_{k}^2$ is not always a positive number (see \cite{SM} for the expression). A special case is that the squared energy $\epsilon_k^{2}$ crosses the negative real axis of the complex plane at critical momentum, i.e., $\epsilon_{k_c}^2 = -C^2$, causing the energy of the positive branch $\epsilon_{k,+}$ to jump from $i|C|$ to $-i|C|$ [Fig.~\ref{Fig2}(b1)]. We regard this as an intrinsic discontinuity point of the energy spectrum. On the one hand, the non-Hermitian energy could turn the second term of Eq.~\eqref{eq3} into a pure real number. On the other hand, near $k = k_c$, the first term is always a real number whether $\epsilon^f_{k,+}$ is a real number or pure imaginary number, the non-Hermitian model may finally lead to a discontinuity in $g_k^{b}(t)$ along the real axis [Fig.~\ref{Fig2}(b2)]. When this happens, $\phi_{k}^{b}(t)$ undergoes a discontinuous change between $\pi$ and $0$, corresponding to a half-integer jump for $\nu^b$.


As a concrete example, we consider a sudden quench from B to A, with the associated dynamics illustrated in Fig.~\ref{Fig2}(c). The DTOP exhibits both half-integer and integer jumps, corresponding to the non-analytical kink of the LR, and also the critical times $t_c$ of the DQPT. This indicates that, DTOP also serves a good criterion for identifying the critical times in non-Hermitian model. 

Meanwhile, in the complex-time plane of Fig.~\ref{Fig2}(d), all $t_c$ points perfectly match the cross-points between the Fisher zeros and the real-time axis (dashed line). In particular, two distinct types of cross-point can be identified: penetrate points (e.g. $t_c = 0.35$) and touch points without crossing (e.g. $t_c = 0.17$). These two cases correspond to integer and half-integer jumps of the DTOP, respectively. According to Eq.~\eqref{eq5}, the energy discontinuity near $k_c$ could separate the left- and right- touch points along the real-time axis and result in broken-like behavior, distinguishing itself from the Hermitian system \cite{SM}.

Furthermore, we plot the momentum-resolved phase $\phi_k^b(t)$ as a function of time in Fig.~\ref{Fig2}(e). 
Four examples of $k$-points are presented to show the contribution to DTOP according to Eq.~\eqref{eq6}. For clear demonstration, we plot the boundary where the discontinuity of $\phi_k^b(t)$ occurs. Here, the solid lines separate the $-\pi$ and $\pi$ phase, corresponding to the red curve in Fig.~\ref{Fig2}(a2). By contrast, the dashed lines separate the $0$ and $\pm \pi$ phases, matching the red curve in Fig.~\ref{Fig2}(b2). Moreover, $t_c$ corresponds to the time coordinate of segment endpoints, which contributes non-zero winding numbers to DTOP. Thus, the dynamics of the DTOP is directly obtained by identifying the positions and types of all segment endpoints, and $\phi_k^b(t)$ diagram brings us more convenience in exploring both the critical times and the momenta. The above results show that DQPTs can be consistently characterized by different approaches, while revealing features unique to non-Hermitian systems.



\begin{figure}[h]
\centering
\includegraphics[width=0.47\textwidth]{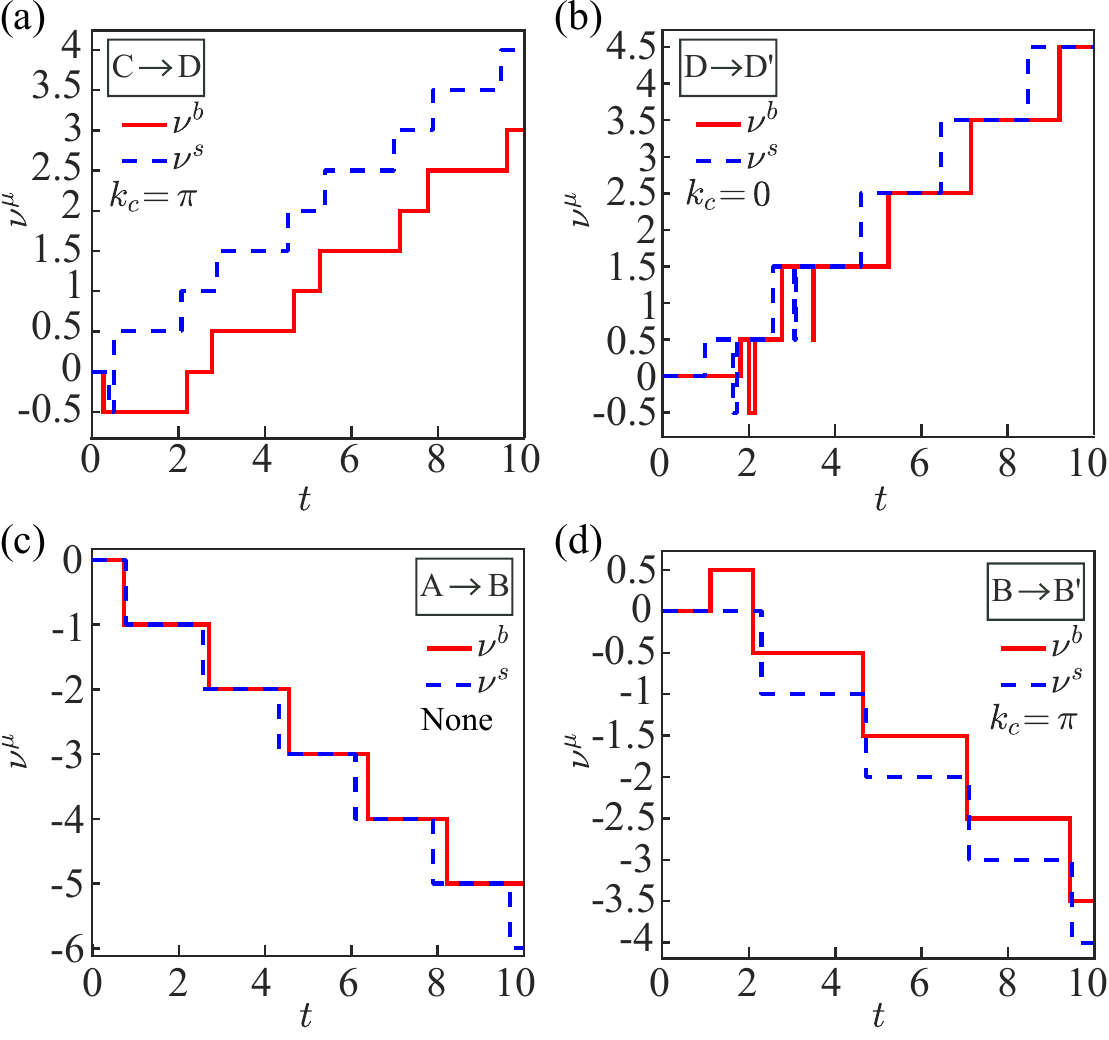}
\caption{\label{Fig3} Dynamical behaviors of the DTOP under different quench protocols, which exhibit periodic (a), single (b) and absence (c) of half-integer jump, respectively. Here, red solid and blue dashed lines denote the DTOP evaluated in the biorthogonal and self-normalized basis, respectively. (d) The case with a single half-jump difference between the two basis.}
\end{figure}

\textcolor{blue}{\it Classification of half-integer DTOPs.--}
Having identified the origin of half-integer jumps associated with non-Hermitian DQPTs, we now proceed to investigate the pattern governing their occurrence. According to the behavior of half-integer jump, the quench processes fall into three categories: (i) periodic half-integer jumps of the DTOP, indicating that the half-integer jumps occur with a fixed period [Fig.~\ref{Fig3}(a)]; (ii) single half-integer jumps [Fig.~\ref{Fig3}(b)], where typically the half-integer jump can only occur once or not; (iii) the complete absence of half-integer jumps [Fig.~\ref{Fig3}(c)]. 

Especially for the biorthogonal basis, we derive a more simplified criterion to observe half-integer jump \cite{SM},

\begin{subequations}
\begin{align}
\mathrm{Im}(\mathbf{d}_{k_c}^i + \mathbf{d}_{k_c}^f)\cdot \mathrm{Re}(\mathbf{d}_{k_c}^i + \mathbf{d}_{k_c}^f) &= 0 \label{eq7} \\
\mathrm{Im}(\mathbf{d}_{k_c}^i)\cdot \mathrm{Re}(\mathbf{d}_{k_c}^i) &= 0 \label{eq8} \\
\mathrm{Im}(\mathbf{d}_{k_c}^f)\cdot \mathrm{Re}(\mathbf{d}_{k_c}^f) &= 0 \label{eq9} \\
|\mathrm{Im}(\mathbf{d}_{k_c}^i)| - |\mathrm{Re}(\mathbf{d}_{k_c}^i)| &> 0 \label{eq10}
\end{align}
\end{subequations}
Eqs.~\eqref{eq7}--\eqref{eq9} describe the orthogonality between the real and imaginary components of the Bloch vectors, whereas Eq.~\eqref{eq10} imposes that $\epsilon^i_{k,+}$ must be discontinuous. We first apply it to the $(\eta,\gamma)$ model, where the critical momenta are located only at $0$ and $\pi$, independently of the specific parameter values. This suggests, as confirmed by the numerical results summarized in Tab. ~\ref{tab:theory}, that the occurrence and classification of half-integer jumps near the critical momentum are entirely determined by the underlying phases of the initial and final states of the quench. On the contrary, in the $(\eta,\delta)$ phase diagram, to guarantee the emergence of half-integer jumps, $\eta$ must remain unchanged after the quench, as the critical momentum explicitly depends on $\eta$. Consequently, even a slight deviation of the quench parameters can lead to the disappearance of a half-integer jump, since Eqs.~\eqref{eq7}--\eqref{eq9} cannot be satisfied at the same time (see Fig. \ref{fig1}(f)).

We further investigate the dependence of half-integer jumps on the phase shown in Tab. ~\ref{tab:theory}. When the complex energy spectra do not cross the imaginary axis, the system remains gapped in the real energy spectrum \cite{SM}, precluding both critical momenta and discontinuous jumps in $\epsilon^i_{k,+}$. Thus, according to Eq.  \eqref{eq10}, quenches from gapped phases, such as I and VI, to any other phase cannot produce half-integer jumps.

\begin{table}[t]
\centering
\caption{\label{tab:theory}
Classification of quench protocols according to half-integer jumps of the DTOP}
\begin{tabular}{ccc}
\toprule
Quench phase & DTOP behavior & Critical momentum\\
\midrule
II $\rightarrow$ I, IV and III $\rightarrow$ IV & Periodic & $\pi$\\	
III $\rightarrow$ II and IV $\rightarrow$ I, II & Periodic & 0\\	
III $\rightarrow$ I & Periodic & 0 and $\pi$\\

II $\rightarrow$ II, III and III $\rightarrow$ II & 1 or 0 & $\pi$ (if exists)\\
III $\rightarrow$ IV and IV $\rightarrow$ III, IV& 1 or 0 & 0  (if exists)\\
III $\rightarrow$ III & 1 or 0 & 0 or $\pi$  (if exists)\\

I $\rightarrow$ Any & 0 & None\\

V $\rightarrow$ VI (same $\eta$) & Periodic & $\cos^{-1}(-\frac{1 + \eta}{1 - \eta})$\\
V $\rightarrow$ V (same $\eta$) & 1 or 0 & $\cos^{-1}(-\frac{1 + \eta}{1 - \eta}) $  (if exists)\\

VI $\rightarrow$ Any & 0 & None\\
V $\rightarrow$ VI, V (different $\eta$) & 0 & None \\
\bottomrule
\end{tabular}
\end{table} 

Building upon Eqs.~\eqref{eq7}--\eqref{eq10}, we further classify the behavior according to the relation between $|\mathrm{Re}(\mathbf{d}_{k_c}^f)|$ and $|\mathrm{Im}(\mathbf{d}_{k_c}^f)|$. If $|\mathrm{Re}(\mathbf{d}_{k_c}^f)| > |\mathrm{Im}(\mathbf{d}_{k_c}^f)|$, $\epsilon^{f}_{k,+}$ is continuous at $k_c$, and  $\mathcal{F}_{k}^\mu$ provides the discontinuity, leading to periodic jumps. Otherwise, each critical momentum contributes at most one half-integer jump. This also implies that periodic jumps typically arise for quenches between different phases, whereas within the same phase at most a single jump can be observed.

Finally, we compare the DTOPs obtained in the two bases for the same quench. Numerically, the self-normalized basis mainly shifts the critical times $t_c$, due to the different jump magnitudes of $\epsilon^f_{k,+}$ and $\mathcal{F}^\mu_k$ near $k_c$. As shown in Fig.~\ref{Fig3}(d), the two bases can differ by a single half-integer jump. Since the self-normalized basis depends explicitly on $|\psi^i_{k,-}\rangle$ [Eq.~\eqref{eq4}], its DTOP is not generally guaranteed to depend solely on the phase. Nevertheless, for the models considered here, it follows the phase-dependent classification summarized in Tab. ~\ref{tab:theory}.

\begin{figure}[h]
\centering
\includegraphics[width=0.47\textwidth]{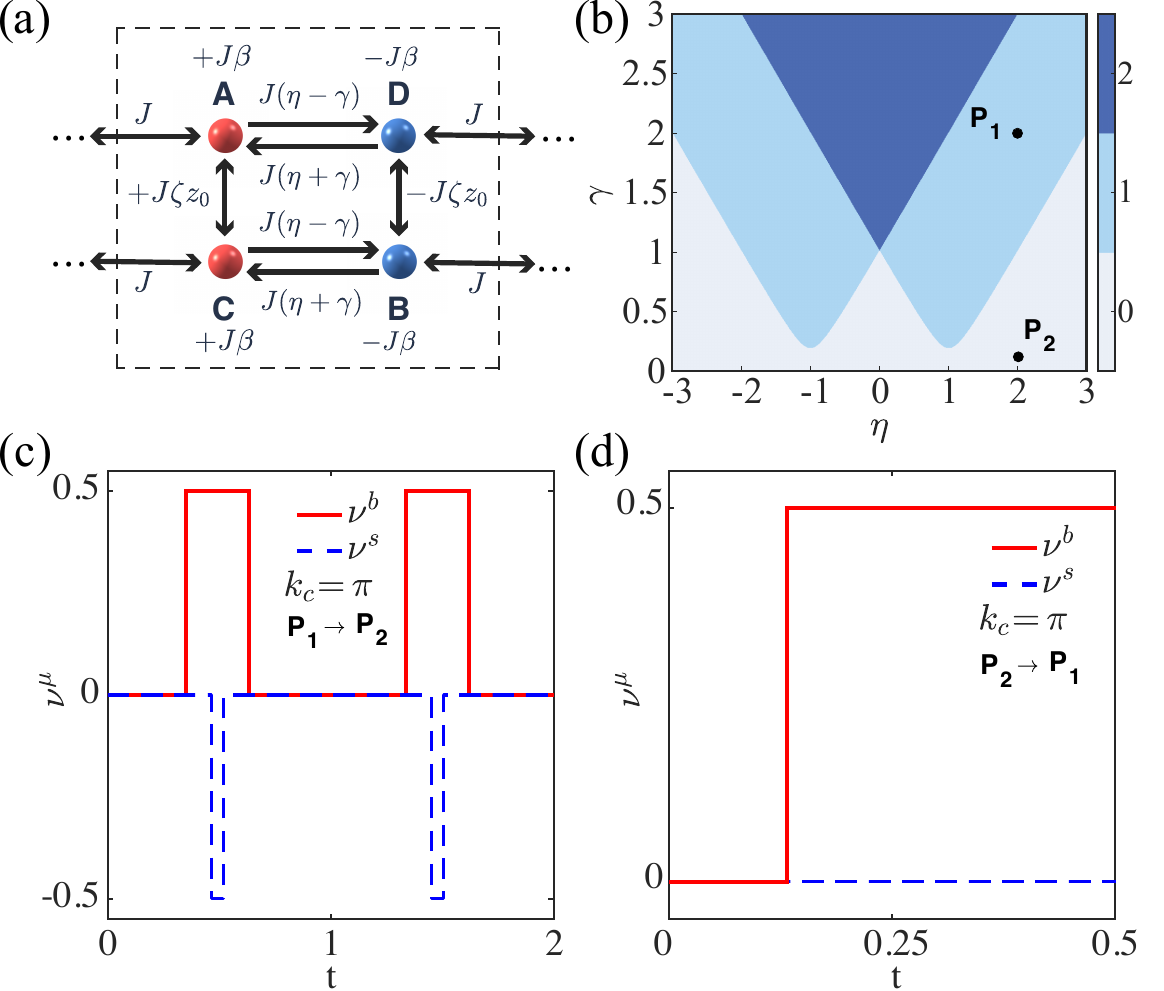}
\caption{\label{Fig4} Fractional dynamical topology in a four-band non-Hermitian chain. (a) Four-orbital lattice with nonreciprocal intracell hoppings $J(\eta\mp\gamma)$, reciprocal intercell hoppings $J$, vertical couplings $\pm J\zeta z_0$, and staggered onsite potentials $\pm J\beta$, where $\beta=\zeta b$. The dashed box denotes a unit cell. (b) Phase diagram in the $(\eta,\gamma)$ plane for the occupied $s=-1$ sector \cite{SM}. The color scale gives the number of complex-energy discontinuities. The coordinates are $P_1=(2,2)$ and $P_2=(2,0.1)$. (c,d) Dynamical topological order parameters for the quenches $P_1\rightarrow P_2$ and $P_2\rightarrow P_1$, respectively. Red solid and blue dashed lines denote the biorthogonal ($\nu^{b}$) and self-normalized ($\nu^{s}$) results. The forward quench exhibits repeated half-integer plateaus, whereas the reverse quench produces a single half-integer jump in $\nu^{b}$ but none in $\nu^{s}$. In both cases, $k_c=\pi$. Other parameters are $J=\pi$, $\zeta=1$, $z_0=0.8$, and $b=1$.}
\end{figure}

\textcolor{blue}{\it Multi-band extension.--} The analytic result of Eq. \eqref{eq3} follows directly from the Cayley–Hamilton theorem \cite{McCarthy1975}, which ensures that any $2\times2$ traceless matrix $H_k$ satisfies $H_k^2\propto I_{2\times2}$. To establish the phase-dependent half-integer jumps in multiband systems, we require that the time-evolution operator to be decomposable into multiple two-dimensional sectors, each having the same form as Eq. \eqref{eq3}. This requires an even number of bands, denoted by $2M$. Defining the projection operators set $\{P_a\}$ where $a = 1, \cdots M$, one obtains $H_{k}^2 = \sum_a \epsilon_{k,a}^2 P_a$ \cite{SM}.

To characterize the correlation between the two independent sectors, except for the three Pauli operators $\Gamma_{a,r} (r=1,2,3)$ within each subspace, all projection operators can be incorporated as additional degrees of freedom of the Hamiltonian. Importantly,  the square of the Hamiltonian still satisfies $H_{k}^2 \in \{ P_a\}$. And the new energy spectrum then takes the universal form $\mathcal{E}_{k,a} = h_a \pm \epsilon_{k,a}$, since the projection operators only induce the energy shift $h_a$. Therefore, the Loschmidt amplitude still admits the following universal form,
\begin{equation}
g_{k}^\mu(t)
=\prod_{q} e^{-i h^f_{a}t}
\left[ \cos(\epsilon_{k,+,q}^f t)
-i\sin(\epsilon_{k,+,q}^f t)
\mathcal{F}_{k,q}^\mu \right].
\label{eq11}
\end{equation}
Taking the four-band case as an example, we construct a real-space model consisting of two coupled SSH chains [Fig.~\ref{Fig4}(a)]. Our numerical results verify this conjecture: the occurrence of $1/2$-jumps is fully determined by the steady-state phase diagram of the system [Fig.~\ref{Fig4}(b)], and either appear periodically [Fig.~\ref{Fig4}(c)] or occur individually [Fig.~\ref{Fig4}(d)] \cite{SM}. 

Based on these results, we further ask whether such phase-diagram-dependent dynamics persists for a generic $2M$-band non-Hermitian Hamiltonian. For convenience, we combine all projectors and introduce a cyclic operator $Q$, satisfying the corresponding cyclic relation
\begin{equation}
    Q = \sum_{a=1}^M e^{2\pi i(a-1)/M} P_a,
\end{equation}
and obviously $Q^M = \mathbb I$. Thus the operators in different subspaces can then be connected through $Q$ and written in a unified form. Defining $\Gamma_0 =\mathbb I$, we find that the Hamiltonian takes the form of Eq. \eqref{eq9}, if and only if it is constructed from the set $\{ Q^a \Gamma_r\}$. Remarkably, this set owns an underlying group structure $\mathbb{Z}_M \times \mathbb{Z}_2 \times \mathbb{Z}_2$. Since it contains only $4M$ elements, whereas a generic $2M$-band Hamiltonian contains $4M^2$ independent matrix components, we conclude that not every multi-band non-Hermitian Hamiltonian can exhibit phase-diagram-dependent half-integer jumps. Such behavior requires the Hamiltonian to belong to the group structure identified above.

\begin{figure}[t]
\centering
\includegraphics[width=0.47\textwidth]{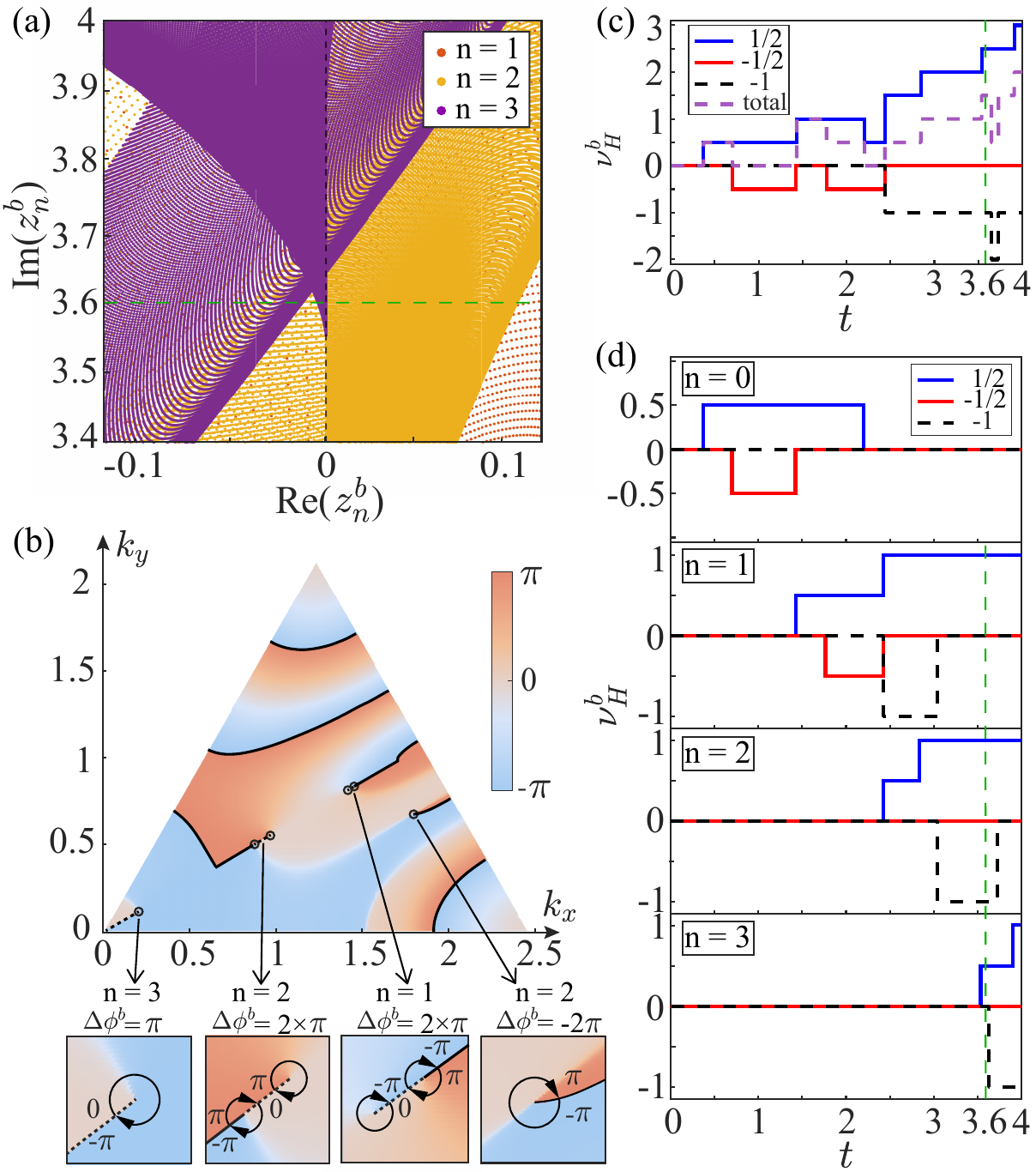}
\caption{\label{Fig5} DQPTs in the non-Hermitian Haldane model formulated within the biorthogonal basis. (a) Fisher zeros diagram for different branches, whose behaviors include crossing the real-time
axis, or touching one side of the axis. (b) The dynamical phase $\phi_{\mathbf{k}}^{b}$ at $t = 3.6$ in the EBZ. The critical momenta are marked with circles. The four zoomed-in panels show the contributions from the endpoints of both dashed and solid lines. Note that each pairs of the dashed lines can merged into a solid line. (c) Total DTOP dynamics of four different types according to the value of phase jump. (d) DTOP dynamics of different branches and types. The half-integer values of DTOP are contributed by the branches $n= 1,2,3$ at $t=3.6$. }
\end{figure}

\textcolor{blue}{\it DQPTs in two-dimensional systems.--} Our approach can be naturally extended to higher dimensions. As an example, we consider DQPTs in the non-Hermitian Haldane model \cite{haldane1988model,li2022gain} and choose quenches across different phases \cite{SM}. In two dimensions, the critical momenta satisfying Eqs.~\eqref{eq7}--\eqref{eq10} form continuous $k_c$-segments rather than isolated points. Correspondingly, the Fisher zeros form planes that either cross or touch the real-time axis [Fig.~\ref{Fig5}(a)] \cite{vajna2015topological,maslowski2024fisher,bhattacharya2017emergent,sacramento2024distribution}. Thus, the nonanalyticities of LR become less informative for identifying phase transitions, while the emergence and disappearance of $k_c$-segments provide a more direct characterization.

We extend the DTOP to two dimensions as \cite{flaschner2018observation}, 
\begin{equation}
\nu_{H}^\mu(t)
= \frac{1}{2\pi}\sum_{\mathbf{k}_{j}} \Delta \phi_{\mathbf{k}_{j}}^\mu(t),
\label{eq11}
\end{equation}
where $\Delta \phi_{\mathbf{k}_{j}}^\mu(t)= \oint_{C_{\mathbf{k}_{j}}} \nabla_{\mathbf{k}} \phi_\mathbf{k}^\mu(t)\cdot d\mathbf{k}$ is the phase winding around a closed loop $C_{\mathbf{k}_{j}}$ enclosing a $\mathbf{k}_c$-segment in the EBZ. By analyzing the phase distribution, we identify multiple $\mathbf{k}_c$-segments [Fig.~\ref{Fig5}(b)]. Segments with a phase difference of $2\pi$ across their two sides contribute integer winding, whereas those with a phase difference of $\pi$ contribute half-integer winding. Contributions from different branches, as well as multiple segments within the same branch, can coexist in two dimensions, leading to overlapping half-integer contributions and, consequently, integer or half-integer total DTOPs [Fig.~\ref{Fig5}(c,d)]

Finally, using Stokes' theorem, the total DTOP can be evaluated directly from the EBZ boundary,
\begin{equation}
    \nu_{H}^\mu(t)=\frac{1}{2\pi}\oint_{\partial_\mathrm{EBZ}} dk\,\partial_k \phi_k^{\mu}(t), 
\end{equation}
so that only the dynamical phase along the EBZ boundary is required to determine the total two-dimensional DTOP.

\textcolor{blue}{\it Experimental feasibility. --} From experimental perspective, the one-dimensional non-Hermitian Hamiltonian can be simulated using a discrete photonic quantum-walk platform, as demonstrated in \cite{Wang2019Simulating}. Here, the two internal states are encoded in photon polarizations, while spatial modes represent lattice sites. The non-unitary dynamics is realized by the polarization-selective loss. In particular, one can reconstruct the momentum-resolved LE and then the corresponding LR, allowing the predicted half-integer jump behavior and its dependence on the phase diagram to be experimentally verified. Moreover, the simulated two-dimensional dynamical vortices in the momentum space could be observed in the ultracold fermionic quantum gases in an optical lattice \cite{flaschner2018observation}.

\textcolor{blue}{\it Discussion.--} In this work, we uncover the physical origin of the half-integer quantization of DTOPs in non-Hermitian systems by studying the even-band model within biorthogonal and self-normalized frameworks. Note that while half-integer jumps have been discovered in Hermitian models, its mechanism in non-Hermitian model is completely different. We demonstrate that such fractional DTOPs are intrinsically determined by the steady-state phase structure of the system Hamiltonian in a specific form, enabling a natural classification of dynamical processes according to the behavior of DTOPs. Our results demonstrate that the half-integer DTOP is a universal feature of non-Hermitian dynamical criticality, rather than a parameter- and model-dependent phenomenon. Furthermore, the theoretical framework developed here can be extended to many-body systems with higher symmetries, offering a versatile tool for characterizing and distinguishing dynamical regimes in interacting open quantum systems. 

\textcolor{blue}{\it Acknowledgments.--} We would like to acknowledge valuable discussions with Xuefeng Zhang and Yaohua Li. This work is supported by NNSFC under Grant No. U25A20198, 12404275, 12574296, and Key Research and Development Program of Shanxi Province (Grants Nos. 202101150101025, 202403021212015). H. S. acknowledges financial support from the Royal Society Newton International Fellowship Alumni follow-on funding (AL201024) of the United Kingdom.

\bibliography{main}

@article{heyl2015scaling,
  title={Scaling and universality at dynamical quantum phase transitions},
  author={Heyl, Markus},
  journal={Physical Review Letters},
  volume={115},
  number={14},
  pages={140602},
  year={2015},
  publisher={APS},
  doi = {10.1103/PhysRevLett.115.140602},
  url = {https://link.aps.org/doi/10.1103/PhysRevLett.115.140602}
}

@article{heyl2013dynamical,
  title={Dynamical quantum phase transitions in the transverse-field Ising model},
  author={Heyl, Markus and Polkovnikov, Anatoli and Kehrein, Stefan},
  journal={Physical Review Letters},
  volume={110},
  number={13},
  pages={135704},
  year={2013},
  publisher={APS},
  doi = {10.1103/PhysRevLett.110.135704},
  url = {https://link.aps.org/doi/10.1103/PhysRevLett.110.135704}
}

@article{jurcevic2017direct,
  title={Direct observation of dynamical quantum phase transitions in an interacting many-body system},
  author={Jurcevic, P and Shen, H and Hauke, P and Maier, C and Brydges, T and Hempel, C and Lanyon, BP and Heyl, Markus and Blatt, R and Roos, CF},
  journal={Physical Review Letters},
  volume={119},
  number={8},
  pages={080501},
  year={2017},
  publisher={APS},
  doi = {10.1103/PhysRevLett.119.080501},
  url = {https://link.aps.org/doi/10.1103/PhysRevLett.119.080501}
}

@article{heyl2018dynamical,
  title={Dynamical quantum phase transitions: a review},
  author={Heyl, Markus},
  journal={Reports on Progress in Physics},
  volume={81},
  number={5},
  pages={054001},
  year={2018},
  publisher={IOP Publishing},
  doi = {10.1088/1361-6633/aaaf9a},
  url = {https://iopscience.iop.org/article/10.1088/1361-6633/aaaf9a}
}

@article{zhang2017observation,
  title={Observation of a many-body dynamical phase transition with a 53-qubit quantum simulator},
  author={Zhang, Jiehang and Pagano, Guido and Hess, Paul W and Kyprianidis, Antonis and Becker, Patrick and Kaplan, Harvey and Gorshkov, Alexey V and Gong, Z-X and Monroe, Christopher},
  journal={Nature},
  volume={551},
  number={7682},
  pages={601--604},
  year={2017},
  publisher={Nature Publishing Group UK London},
  doi = {10.1038/nature24654},
  url = {https://doi.org/10.1038/nature24654}

}

@article{budich2016dynamical,
  title={Dynamical topological order parameters far from equilibrium},
  author={Budich, Jan Carl and Heyl, Markus},
  journal={Physical Review B },
  volume={93},
  number={8},
  pages={085416},
  year={2016},
  publisher={APS},
  doi = {10.1103/PhysRevB.93.085416},
  url = {https://link.aps.org/doi/10.1103/PhysRevB.93.085416}
}

@article{wang2019simulating,
  title={Simulating dynamic quantum phase transitions in photonic quantum walks},
  author={Wang, Kunkun and Qiu, Xingze and Xiao, Lei and Zhan, Xiang and Bian, Zhihao and Yi, Wei and Xue, Peng},
  journal={Physical Review Letters},
  volume={122},
  number={2},
  pages={020501},
  year={2019},
  publisher={APS},
  doi = {10.1103/PhysRevLett.122.020501},
  url = {https://link.aps.org/doi/10.1103/PhysRevLett.122.020501}
}

@article{xu2020measuring,
  title={Measuring a dynamical topological order parameter in quantum walks},
  author={Xu, Xiao-Ye and Wang, Qin-Qin and Heyl, Markus and Budich, Jan Carl and Pan, Wei-Wei and Chen, Zhe and Jan, Munsif and Sun, Kai and Xu, Jin-Shi and Han, Yong-Jian and others},
  journal={Light: Science \& Applications},
  volume={9},
  number={1},
  pages={7},
  year={2020},
  publisher={Nature Publishing Group UK London},
  doi = {10.1038/s41377-019-0237-8},
  url = {https://doi.org/10.1038/s41377-019-0237-8}
}

@article{yang2019floquet,
  title={Floquet dynamical quantum phase transitions},
  author={Yang, Kai and Zhou, Longwen and Ma, Wenchao and Kong, Xi and Wang, Pengfei and Qin, Xi and Rong, Xing and Wang, Ya and Shi, Fazhan and Gong, Jiangbin and others},
  journal={Physical Review B },
  volume={100},
  number={8},
  pages={085308},
  year={2019},
  publisher={APS},
  doi = {10.1103/PhysRevB.100.085308},
  url = {https://link.aps.org/doi/10.1103/PhysRevB.100.085308}
}

@article{naji2022dissipative,
  title={Dissipative floquet dynamical quantum phase transition},
  author={Naji, J and Jafari, Masoud and Jafari, R and Akbari, Alireza},
  journal={Physical Review A},
  volume={105},
  number={2},
  pages={022220},
  year={2022},
  publisher={APS},
  doi = {10.1103/PhysRevA.105.022220},
  url = {https://link.aps.org/doi/10.1103/PhysRevA.105.022220}
}

@article{jafari2022floquet,
  title={Floquet dynamical quantum phase transitions under synchronized periodic driving},
  author={Jafari, R and Akbari, Alireza and Mishra, Utkarsh and Johannesson, Henrik},
  journal={Physical Review B },
  volume={105},
  number={9},
  pages={094311},
  year={2022},
  publisher={APS},
  doi = {10.1103/PhysRevB.105.094311},
  url = {https://link.aps.org/doi/10.1103/PhysRevB.105.094311}
}

@article{jafari2021floquet,
  title={Floquet dynamical phase transition and entanglement spectrum},
  author={Jafari, R and Akbari, Alireza},
  journal={Physical Review A},
  volume={103},
  number={1},
  pages={012204},
  year={2021},
  publisher={APS},
  doi = {10.1103/PhysRevA.103.012204},
  url = {https://link.aps.org/doi/10.1103/PhysRevA.103.012204}
}

@article{sharma2016slow,
  title={Slow quenches in a quantum Ising chain: Dynamical phase transitions and topology},
  author={Sharma, Shraddha and Divakaran, Uma and Polkovnikov, Anatoli and Dutta, Amit},
  journal={Physical Review B },
  volume={93},
  number={14},
  pages={144306},
  year={2016},
  publisher={APS},
  doi = {10.1103/PhysRevB.93.144306},
  url = {https://link.aps.org/doi/10.1103/PhysRevB.93.144306}
}

@article{yokomizo2019non,
  title={Non-Bloch band theory of non-Hermitian systems},
  author={Yokomizo, Kazuki and Murakami, Shuichi},
  journal={Physical Review Letters},
  volume={123},
  number={6},
  pages={066404},
  year={2019},
  publisher={APS},
  doi = {10.1103/PhysRevLett.123.066404},
  url = {https://link.aps.org/doi/10.1103/PhysRevLett.123.066404}
}

@article{yao2018edge,
  title={Edge states and topological invariants of non-Hermitian systems},
  author={Yao, Shunyu and Wang, Zhong},
  journal={Physical Review Letters},
  volume={121},
  number={8},
  pages={086803},
  year={2018},
  publisher={APS},
  doi = {10.1103/PhysRevLett.121.086803},
  url = {https://link.aps.org/doi/10.1103/PhysRevLett.121.086803}
}

@article{yokomizo2020non,
  title={Non-Bloch band theory and bulk--edge correspondence in non-Hermitian systems},
  author={Yokomizo, Kazuki and Murakami, Shuichi},
  journal={Progress of Theoretical and Experimental Physics},
  volume={2020},
  number={12},
  pages={12A102},
  year={2020},
  publisher={Oxford University Press},
  doi = {10.1093/ptep/ptaa140},
  url = {https://doi.org/10.1093/ptep/ptaa140}
}

@article{shen2018topological,
  title={Topological band theory for non-Hermitian Hamiltonians},
  author={Shen, Huitao and Zhen, Bo and Fu, Liang},
  journal={Physical Review Letters},
  volume={120},
  number={14},
  pages={146402},
  year={2018},
  publisher={APS},
  doi = {10.1103/PhysRevLett.120.146402},
  url = {https://link.aps.org/doi/10.1103/PhysRevLett.120.146402}
}

@article{ghatak2019new,
  title={New topological invariants in non-Hermitian systems},
  author={Ghatak, Ananya and Das, Tanmoy},
  journal={Journal of Physics: Condensed Matter},
  volume={31},
  number={26},
  pages={263001},
  year={2019},
  publisher={IOP Publishing},
  doi = {10.1088/1361-648X/ab11b3},
  url = {https://iopscience.iop.org/article/10.1088/1361-648X/ab11b3}
}

@article{yin2018geometrical,
  title={Geometrical meaning of winding number and its characterization of topological phases in one-dimensional chiral non-Hermitian systems},
  author={Yin, Chuanhao and Jiang, Hui and Li, Linhu and L{\"u}, Rong and Chen, Shu},
  journal={Physical Review A},
  volume={97},
  number={5},
  pages={052115},
  year={2018},
  publisher={APS},
  doi = {10.1103/PhysRevA.97.052115},
  url = {https://link.aps.org/doi/10.1103/PhysRevA.97.052115}

}

@article{gong2018topological,
  title={Topological phases of non-Hermitian systems},
  author={Gong, Zongping and Ashida, Yuto and Kawabata, Kohei and Takasan, Kazuaki and Higashikawa, Sho and Ueda, Masahito},
  journal={Physical Review X},
  volume={8},
  number={3},
  pages={031079},
  year={2018},
  publisher={APS},
  doi = {10.1103/PhysRevX.8.031079},
  url = {https://link.aps.org/doi/10.1103/PhysRevX.8.031079}
}

@article{kawabata2019symmetry,
  title={Symmetry and topology in non-Hermitian physics},
  author={Kawabata, Kohei and Shiozaki, Ken and Ueda, Masahito and Sato, Masatoshi},
  journal={Physical Review X},
  volume={9},
  number={4},
  pages={041015},
  year={2019},
  publisher={APS},
  doi = {10.1103/PhysRevX.9.041015},
  url = {https://link.aps.org/doi/10.1103/PhysRevX.9.041015}

}

@article{wang2021topological,
  title={Topological complex-energy braiding of non-Hermitian bands},
  author={Wang, Kai and Dutt, Avik and Wojcik, Charles C and Fan, Shanhui},
  journal={Nature},
  volume={598},
  number={7879},
  pages={59--64},
  year={2021},
  publisher={Nature Publishing Group UK London},
  doi = {10.1038/s41586-021-03848-x},
  url = {https://doi.org/10.1038/s41586-021-03848-x}
}

@article{bergholtz2021exceptional,
  title={Exceptional topology of non-Hermitian systems},
  author={Bergholtz, Emil J and Budich, Jan Carl and Kunst, Flore K},
  journal={Reviews of Modern Physics},
  volume={93},
  number={1},
  pages={015005},
  year={2021},
  publisher={APS},
  doi = {10.1103/RevModPhys.93.015005},
  url = {https://link.aps.org/doi/10.1103/RevModPhys.93.015005}
}

@article{ding2022non,
  title={Non-Hermitian topology and exceptional-point geometries},
  author={Ding, Kun and Fang, Chen and Ma, Guancong},
  journal={Nature Reviews Physics},
  volume={4},
  number={12},
  pages={745--760},
  year={2022},
  publisher={Nature Publishing Group UK London},
  url = {https://doi.org/10.1038/s42254-022-00516-5},
  doi = {10.1038/s42254-022-00516-5}
}

@article{bender1998real,
  title={Real Spectra in Non-Hermitian Hamiltonians Having PT Symmetry},
  author={Bender, Carl M and Boettcher, Stefan},
  journal={Physical Review Letters},
  volume={80},
  number={24},
  pages={5243},
  year={1998},
  publisher={APS},
  doi = {10.1103/PhysRevLett.80.5243},
  url = {https://link.aps.org/doi/10.1103/PhysRevLett.80.5243}
}

@article{helbig2020generalized,
  title={Generalized bulk--boundary correspondence in non-Hermitian topolectrical circuits},
  author={Helbig, Tobias and Hofmann, Tobias and Imhof, Sven and Abdelghany, M and Kiessling, Tobias and Molenkamp, Laurens W and Lee, Ching Hua and Szameit, Alexander and Greiter, Martin and Thomale, Ronny},
  journal={Nature Physics},
  volume={16},
  number={7},
  pages={747--750},
  year={2020},
  publisher={Nature Publishing Group UK London},
  doi = {10.1038/s41567-020-0922-9},
  url = {https://doi.org/10.1038/s41567-020-0922-9}
}

@article{zhang2021observation,
  title={Observation of higher-order non-Hermitian skin effect},
  author={Zhang, Xiujuan and Tian, Yuan and Jiang, Jian-Hua and Lu, Ming-Hui and Chen, Yan-Feng},
  journal={Nature Communications},
  volume={12},
  number={1},
  pages={5377},
  year={2021},
  publisher={Nature Publishing Group UK London},
  doi = {10.1038/s41467-021-25716-y},
  url = {https://doi.org/10.1038/s41467-021-25716-y}
}

@article{Zhang2020Correspondence,
  title = {Correspondence between Winding Numbers and Skin Modes in Non-Hermitian Systems},
  author = {Zhang, Kai and Yang, Zhesen and Fang, Chen},
  journal = {Physical Review Letters},
  volume = {125},
  issue = {12},
  pages = {126402},
  numpages = {6},
  year = {2020},
  month = {Sep},
  publisher = {American Physical Society},
  doi = {10.1103/PhysRevLett.125.126402},
  url = {https://link.aps.org/doi/10.1103/PhysRevLett.125.126402}
}

@article{PhysRevLett.121.026808,
  title = {Biorthogonal Bulk-Boundary Correspondence in Non-Hermitian Systems},
  author = {Kunst, Flore K. and Edvardsson, Elisabet and Budich, Jan Carl and Bergholtz, Emil J.},
  journal = {Physical Review Letters},
  volume = {121},
  issue = {2},
  pages = {026808},
  numpages = {6},
  year = {2018},
  month = {Jul},
  publisher = {American Physical Society},
  doi = {10.1103/PhysRevLett.121.026808},
  url = {https://link.aps.org/doi/10.1103/PhysRevLett.121.026808}
}

@article{edvardsson2019non,
  title={Non-Hermitian extensions of higher-order topological phases and their biorthogonal bulk-boundary correspondence},
  author={Edvardsson, Elisabet and Kunst, Flore K and Bergholtz, Emil J},
  journal={Physical Review B },
  volume={99},
  number={8},
  pages={081302},
  year={2019},
  publisher={APS},
  doi = {10.1103/PhysRevB.99.081302},
  url = {https://link.aps.org/doi/10.1103/PhysRevB.99.081302}
}

@article{edvardsson2020phase,
  title={Phase transitions and generalized biorthogonal polarization in non-Hermitian systems},
  author={Edvardsson, Elisabet and Kunst, Flore K and Yoshida, Tsuneya and Bergholtz, Emil J},
  journal={Physical Review Research},
  volume={2},
  number={4},
  pages={043046},
  year={2020},
  publisher={APS},
  doi = {10.1103/PhysRevResearch.2.043046},
  url = {https://link.aps.org/doi/10.1103/PhysRevResearch.2.043046}
}

@article{tang2022dynamical,
  title={Dynamical scaling of Loschmidt echo in non-Hermitian systems},
  author={Tang, Jia-Chen and Kou, Su-Peng and Sun, Gaoyong},
  journal={Europhysics Letters},
  volume={137},
  number={4},
  pages={40001},
  year={2022},
  publisher={EDP Sciences, IOP Publishing and Societ{\`a} Italiana di Fisica},
  doi = {10.1209/0295-5075/ac53c4},
  url = {https://iopscience.iop.org/article/10.1209/0295-5075/ac53c4}
}

@article{sun2022biorthogonal,
  title={Biorthogonal quantum criticality in non-Hermitian many-body systems},
  author={Sun, Gaoyong and Tang, Jia-Chen and Kou, Su-Peng},
  journal={Frontiers of Physics},
  volume={17},
  number={3},
  pages={33502},
  year={2022},
  publisher={Springer},
  doi = {10.1007/s11467-021-1126-1},
  url = {https://doi.org/10.1007/s11467-021-1126-1}
}

@article{mondal2022anomaly,
  title={Anomaly in the dynamical quantum phase transition in a non-Hermitian system with extended gapless phases},
  author={Mondal, Debashish and Nag, Tanay},
  journal={Physical Review B },
  volume={106},
  number={5},
  pages={054308},
  year={2022},
  publisher={APS},
  doi = {10.1103/PhysRevB.106.054308},
  url = {https://link.aps.org/doi/10.1103/PhysRevB.106.054308}
}

@article{jing2024biorthogonal,
  title={Biorthogonal dynamical quantum phase transitions in non-Hermitian systems},
  author={Jing, Yecheng and Dong, Jian-Jun and Zhang, Yu-Yu and Hu, Zi-Xiang},
  journal={Physical Review Letters},
  volume={132},
  number={22},
  pages={220402},
  year={2024},
  publisher={APS},
  doi = {10.1103/PhysRevLett.132.220402},
  url = {https://link.aps.org/doi/10.1103/PhysRevLett.132.220402}
}

@article{su1980soliton,
  title={Soliton excitations in polyacetylene},
  author={Su, W-P\_ and Schrieffer, JR and Heeger, AJ},
  journal={Physical Review B },
  volume={22},
  number={4},
  pages={2099},
  year={1980},
  publisher={APS},
  doi = {10.1103/PhysRevB.22.2099},
  url = {https://link.aps.org/doi/10.1103/PhysRevB.22.2099}
}

@article{brody2014biorthogonal,
  title={Biorthogonal quantum mechanics},
  author={Brody, Dorje C},
  journal={Journal of Physics A: Mathematical and Theoretical},
  volume={47},
  number={3},
  pages={035305},
  year={2014},
  publisher={IOP Publishing},
  doi = {10.1088/1751-8113/47/3/035305},
  url = {https://iopscience.iop.org/article/10.1088/1751-8113/47/3/035305}
}

@article{zhou2018dynamical,
  title={Dynamical quantum phase transitions in non-Hermitian lattices},
  author={Zhou, Longwen and Wang, Qing-hai and Wang, Hailong and Gong, Jiangbin},
  journal={Physical Review A},
  volume={98},
  number={2},
  pages={022129},
  year={2018},
  publisher={APS},
  doi = {10.1103/PhysRevA.98.022129},
  url = {https://link.aps.org/doi/10.1103/PhysRevA.98.022129}
}

@article{gong2018piecewise,
  title={Piecewise adiabatic following in non-Hermitian cycling},
  author={Gong, Jiangbin and Wang, Qing-hai},
  journal={Physical Review A},
  volume={97},
  number={5},
  pages={052126},
  year={2018},
  publisher={APS},
  doi = {10.1103/PhysRevA.97.052126},
  url = {https://link.aps.org/doi/10.1103/PhysRevA.97.052126}
}

@article{aharonov1987phase,
  title={Phase change during a cyclic quantum evolution},
  author={Aharonov, Yakir and Anandan, J},
  journal={Physical Review Letters},
  volume={58},
  number={16},
  pages={1593},
  year={1987},
  publisher={APS},
  doi = {10.1103/PhysRevLett.58.1593},
  url = {https://link.aps.org/doi/10.1103/PhysRevLett.58.1593}
}

@article{li2022gain,
  title={Gain-loss-induced hybrid skin-topological effect},
  author={Li, Yaohua and Liang, Chao and Wang, Chenyang and Lu, Cuicui and Liu, Yong-Chun},
  journal={Physical Review Letters},
  volume={128},
  number={22},
  pages={223903},
  year={2022},
  publisher={APS},
  doi = {10.1103/PhysRevLett.128.223903},
  url = {https://link.aps.org/doi/10.1103/PhysRevLett.128.223903}
}

@article{haldane1988model,
  title={Model for a Quantum Hall Effect without Landau Levels: Condensed-Matter Realization of the "Parity Anomaly"},
  author={Haldane, F Duncan M},
  journal={Physical Review Letters},
  volume={61},
  number={18},
  pages={2015},
  year={1988},
  publisher={APS},
  doi = {10.1103/PhysRevLett.61.2015},
  url = {https://link.aps.org/doi/10.1103/PhysRevLett.61.2015}
}

@article{vajna2015topological,
  title={Topological classification of dynamical phase transitions},
  author={Vajna, Szabolcs and D{\'o}ra, Bal{\'a}zs},
  journal={Physical Review B },
  volume={91},
  number={15},
  pages={155127},
  year={2015},
  publisher={APS},
  doi = {10.1103/PhysRevB.91.155127},
  url = {https://link.aps.org/doi/10.1103/PhysRevB.91.155127}
}

@article{maslowski2024fisher,
  title={Fisher zeroes and dynamical quantum phase transitions for two-and three-dimensional models},
  author={Mas{\l}owski, Tomasz and Cheraghi, Hadi and Sirker, Jesko and Sedlmayr, Nicholas},
  journal={Physical Review B },
  volume={110},
  number={22},
  pages={224302},
  year={2024},
  publisher={APS},
  doi = {10.1103/PhysRevB.110.224302},
  url = {https://link.aps.org/doi/10.1103/PhysRevB.110.224302}
}

@article{bhattacharya2017emergent,
  title={Emergent topology and dynamical quantum phase transitions in two-dimensional closed quantum systems},
  author={Bhattacharya, Utso and Dutta, Amit},
  journal={Physical Review B },
  volume={96},
  number={1},
  pages={014302},
  year={2017},
  publisher={APS},
  doi = {10.1103/PhysRevB.96.014302},
  url = {https://link.aps.org/doi/10.1103/PhysRevB.96.014302}
}

@article{sacramento2024distribution,
  title={Distribution of Fisher zeros in dynamical quantum phase transitions of two-dimensional topological systems},
  author={Sacramento, PD and Yu, Wing Chi},
  journal={Physical Review B },
  volume={109},
  number={13},
  pages={134301},
  year={2024},
  publisher={APS},
  doi = {10.1103/PhysRevB.109.134301},
  url = {https://link.aps.org/doi/10.1103/PhysRevB.109.134301}
}

@article{flaschner2018observation,
  title={Observation of dynamical vortices after quenches in a system with topology},
  author={Fl{\"a}schner, Nick and Vogel, Dominik and Tarnowski, Matthias and Rem, Benno S and L{\"u}hmann, D-S and Heyl, Markus and Budich, Jan Carl and Mathey, Ludwig and Sengstock, Klaus and Weitenberg, Christof},
  journal={Nature Physics},
  volume={14},
  number={3},
  pages={265--268},
  year={2018},
  publisher={Nature Publishing Group UK London},
  doi = {10.1038/s41567-017-0013-8},
  url = {https://doi.org/10.1038/s41567-017-0013-8}
}

@article{Tian2019Observation,
  title = {Observation of dynamical phase transitions in a topological nanomechanical system},
  author = {Tian, Tian and Ke, Yongguan and Zhang, Liang and Lin, Shaochun and Shi, Zhifu and Huang, Pu and Lee, Chaohong and Du, Jiangfeng},
  journal = {Physical Review B},
  volume = {100},
  issue = {2},
  pages = {024310},
  numpages = {12},
  year = {2019},
  month = {Jul},
  publisher = {American Physical Society},
  doi = {10.1103/PhysRevB.100.024310},
  url = {https://link.aps.org/doi/10.1103/PhysRevB.100.024310}
}

@article{Non-Hermitian2020Yuto,
author = {Yuto Ashida and Zongping Gong and Masahito Ueda},
title = {Non-Hermitian physics},
journal = {Advances in Physics},
volume = {69},
number = {3},
pages = {249--435},
year = {2020},
publisher = {Taylor \& Francis},
doi = {10.1080/00018732.2021.1876991},
URL = {  https://doi.org/10.1080/00018732.2021.1876991},
}

@article{Xiao2020Non-Hermitian,
   author = {Xiao, Lei and Deng, Tianshu and Wang, Kunkun and Zhu, Gaoyan and Wang, Zhong and Yi, Wei and Xue, Peng},
   title = {Non-Hermitian bulk–boundary correspondence in quantum dynamics},
   journal = {Nature Physics},
   volume = {16},
   number = {7},
   pages = {761-766},
   ISSN = {1745-2481},
   DOI = {10.1038/s41567-020-0836-6},
   url = {https://doi.org/10.1038/s41567-020-0836-6},
   year = {2020},
   type = {Journal Article}
}

@article{Lang2018Dynamical,
  title = {Dynamical quantum phase transition for mixed states in open systems},
  author = {Lang, Haifeng and Chen, Yixin and Hong, Qiantan and Fan, Heng},
  journal = {Physical Review B},
  volume = {98},
  issue = {13},
  pages = {134310},
  numpages = {7},
  year = {2018},
  month = {Oct},
  publisher = {American Physical Society},
  doi = {10.1103/PhysRevB.98.134310},
  url = {https://link.aps.org/doi/10.1103/PhysRevB.98.134310}
}

@article{Kai2020Probing,
author = {Kai Xu  and Zheng-Hang Sun  and Wuxin Liu  and Yu-Ran Zhang  and Hekang Li  and Hang Dong  and Wenhui Ren  and Pengfei Zhang  and Franco Nori  and Dongning Zheng  and Heng Fan  and H. Wang },
title = {Probing dynamical phase transitions with a superconducting quantum simulator},
journal = {Science Advances},
volume = {6},
number = {25},
pages = {eaba4935},
year = {2020},
doi = {10.1126/sciadv.aba4935},
URL = {https://www.science.org/doi/abs/10.1126/sciadv.aba4935},
}

@article{Zeng2023Dynamical,
  title = {Dynamical singularity of the rate function for quench dynamics in finite-size quantum systems},
  author = {Zeng, Yumeng and Zhou, Bozhen and Chen, Shu},
  journal = {Physical Review B},
  volume = {107},
  issue = {13},
  pages = {134302},
  numpages = {10},
  year = {2023},
  month = {Apr},
  publisher = {American Physical Society},
  doi = {10.1103/PhysRevB.107.134302},
  url = {https://link.aps.org/doi/10.1103/PhysRevB.107.134302}
}

@article{Nie2020Experimental,
  title = {Experimental Observation of Equilibrium and Dynamical Quantum Phase Transitions via Out-of-Time-Ordered Correlators},
  author = {Nie, Xinfang and Wei, Bo-Bo and Chen, Xi and Zhang, Ze and Zhao, Xiuzhu and Qiu, Chudan and Tian, Yu and Ji, Yunlan and Xin, Tao and Lu, Dawei and Li, Jun},
  journal = {Physical Review Letters},
  volume = {124},
  issue = {25},
  pages = {250601},
  numpages = {6},
  year = {2020},
  month = {Jun},
  publisher = {American Physical Society},
  doi = {10.1103/PhysRevLett.124.250601},
  url = {https://link.aps.org/doi/10.1103/PhysRevLett.124.250601}
}

@article{Zhou2021Non-Hermitian,
doi = {10.1088/1367-2630/ac0574},
url = {https://doi.org/10.1088/1367-2630/ac0574},
year = {2021},
month = {jun},
publisher = {IOP Publishing},
volume = {23},
number = {6},
pages = {063041},
author = {Zhou, Longwen and Du, Qianqian},
title = {Non-Hermitian topological phases and dynamical quantum phase transitions: a generic connection},
journal = {New Journal of Physics},
}

@article{Mondal2023Finite,
  title = {Finite-temperature dynamical quantum phase transition in a non-Hermitian system},
  author = {Mondal, Debashish and Nag, Tanay},
  journal = {Physical Review B},
  volume = {107},
  issue = {18},
  pages = {184311},
  numpages = {14},
  year = {2023},
  month = {May},
  publisher = {American Physical Society},
  doi = {10.1103/PhysRevB.107.184311},
  url = {https://link.aps.org/doi/10.1103/PhysRevB.107.184311}
}

@article{Bhattacharya2017Mixed,
  title = {Mixed state dynamical quantum phase transitions},
  author = {Bhattacharya, Utso and Bandyopadhyay, Souvik and Dutta, Amit},
  journal = {Physical Review B},
  volume = {96},
  issue = {18},
  pages = {180303},
  numpages = {5},
  year = {2017},
  month = {Nov},
  publisher = {American Physical Society},
  doi = {10.1103/PhysRevB.96.180303},
  url = {https://link.aps.org/doi/10.1103/PhysRevB.96.180303}
}

@article{Zhang2025Self-normal,
   author = {Zhang, Haiting and Wang, Kunkun and Xiao, Lei and Xue, Peng},
   title = {Self-normal and biorthogonal dynamical quantum phase transitions in non-Hermitian quantum walks},
   journal = {Light: Science \& Applications},
   volume = {14},
   number = {1},
   pages = {253},
   ISSN = {2047-7538},
   DOI = {10.1038/s41377-025-01919-6},
   url = {https://doi.org/10.1038/s41377-025-01919-6},
   year = {2025},
   type = {Journal Article}
}

@article{Peotta2021Determination,
  title = {Determination of Dynamical Quantum Phase Transitions in Strongly Correlated Many-Body Systems Using Loschmidt Cumulants},
  author = {Peotta, Sebastiano and Brange, Fredrik and Deger, Aydin and Ojanen, Teemu and Flindt, Christian},
  journal = {Physical Review X},
  volume = {11},
  issue = {4},
  pages = {041018},
  numpages = {15},
  year = {2021},
  month = {Oct},
  publisher = {American Physical Society},
  doi = {10.1103/PhysRevX.11.041018},
  url = {https://link.aps.org/doi/10.1103/PhysRevX.11.041018}
}

@article{Guo2019Observation,
  title = {Observation of a Dynamical Quantum Phase Transition by a Superconducting Qubit Simulation},
  author = {Guo, Xue-Yi and Yang, Chao and Zeng, Yu and Peng, Yi and Li, He-Kang and Deng, Hui and Jin, Yi-Rong and Chen, Shu and Zheng, Dongning and Fan, Heng},
  journal = {Physical Review Applied},
  volume = {11},
  issue = {4},
  pages = {044080},
  numpages = {12},
  year = {2019},
  month = {Apr},
  publisher = {American Physical Society},
  doi = {10.1103/PhysRevApplied.11.044080},
  url = {https://link.aps.org/doi/10.1103/PhysRevApplied.11.044080}
}

@article{Sadrzadeh2021Dynamical,
  title = {Dynamical topological quantum phase transitions at criticality},
  author = {Sadrzadeh, M. and Jafari, R. and Langari, A.},
  journal = {Physical Review B},
  volume = {103},
  issue = {14},
  pages = {144305},
  numpages = {10},
  year = {2021},
  month = {Apr},
  publisher = {American Physical Society},
  doi = {10.1103/PhysRevB.103.144305},
  url = {https://link.aps.org/doi/10.1103/PhysRevB.103.144305}
}

@article{Yu2021Correlations,
  title = {Correlations and dynamical quantum phase transitions in an interacting topological insulator},
  author = {Yu, Wing Chi and Sacramento, P. D. and Li, Yan Chao and Lin, Hai-Qing},
  journal = {Physical Review B},
  volume = {104},
  issue = {8},
  pages = {085104},
  numpages = {10},
  year = {2021},
  month = {Aug},
  publisher = {American Physical Society},
  doi = {10.1103/PhysRevB.104.085104},
  url = {https://link.aps.org/doi/10.1103/PhysRevB.104.085104}
}

@article{Heyl2017Quenching,
  title = {Quenching a quantum critical state by the order parameter: Dynamical quantum phase transitions and quantum speed limits},
  author = {Heyl, Markus},
  journal = {Physical Review B},
  volume = {95},
  issue = {6},
  pages = {060504},
  numpages = {5},
  year = {2017},
  month = {Feb},
  publisher = {American Physical Society},
  doi = {10.1103/PhysRevB.95.060504},
  url = {https://link.aps.org/doi/10.1103/PhysRevB.95.060504}
}

@article{Chen2020Experimentally,
  title = {Experimentally detecting dynamical quantum phase transitions in a slowly quenched Ising-chain model},
  author = {Chen, Zhe and Cui, Jin-Ming and Ai, Ming-Zhong and He, Ran and Huang, Yun-Feng and Han, Yong-Jian and Li, Chuan-Feng and Guo, Guang-Can},
  journal = {Physical Review A},
  volume = {102},
  issue = {4},
  pages = {042222},
  numpages = {9},
  year = {2020},
  month = {Oct},
  publisher = {American Physical Society},
  doi = {10.1103/PhysRevA.102.042222},
  url = {https://link.aps.org/doi/10.1103/PhysRevA.102.042222}
}

@article{ZHANG2026131193,
title = {Phase quench effects on dynamical quantum phase transitions in the Su-Schrieffer-Heeger model},
journal = {Physics Letters A},
volume = {566},
pages = {131193},
year = {2026},
issn = {0375-9601},
doi = {https://doi.org/10.1016/j.physleta.2025.131193},
url = {https://www.sciencedirect.com/science/article/pii/S0375960125009739},
author = {Mei-Yu Zhang and Wen-Hui Nie and Lin-Cheng Wang}}

@article{Xu_2026,
doi = {10.1088/1367-2630/ae635c},
url = {https://doi.org/10.1088/1367-2630/ae635c},
year = {2026},
month = {may},
publisher = {IOP Publishing},
volume = {28},
number = {5},
pages = {054505},
author = {Xu, Bao-Ming},
title = {Topological classification of dynamical quantum phase transitions in the 1D XY model via critical mode analysis},
journal = {New Journal of Physics}
}

@article{McCarthy1975,
  author  = {C. A. McCarthy},
  title   = {The Cayley--Hamilton Theorem},
  journal = {The American Mathematical Monthly},
  volume  = {82},
  pages   = {390--391},
  year    = {1975},
  doi     = {10.1080/00029890.1975.11993841}
}

@article{SM,
  author  = {1},
  title={See Supplemental Material for the phase diagram; the derivation on both basis; the exact expression on DTOP; the derivations on the conditions for the occurence of the half-integer DTOPs; The details in the non-Hermitian Haldane model; The universal theory for constructing the multiband Hamiltonian},
  journal = {1},
  volume  = {1},
  pages   = {1},
  year    = {1},
  doi     = {1}
}
\end{document}